\documentclass[preprint,12pt]{elsarticle}

\usepackage{amssymb}

\usepackage{lineno}

\usepackage{color}
\usepackage{amsmath}
\usepackage{txfonts}
\usepackage{url}
 \usepackage{tabularx}
 \usepackage{ulem}
  \usepackage{lscape}

 \definecolor{mygreen}{rgb}{0,0.4,0}

\journal{Physics Letters B}

\begin{document}

\begin{frontmatter}



\title{Search for dark matter in the form of hidden photons and axion-like particles in the XMASS detector }

\author[ICRR,IPMU]{K.~Abe}
\author[ICRR,IPMU]{K.~Hiraide}
\author[ICRR,IPMU]{K.~Ichimura}
\author[ICRR,IPMU]{Y.~Kishimoto}
\author[ICRR,IPMU]{K.~Kobayashi}
\author[ICRR]{M.~Kobayashi}
\author[ICRR,IPMU]{S.~Moriyama}
\author[ICRR,IPMU]{M.~Nakahata}
\author[ICRR,IPMU]{H.~Ogawa\fnref{OgawaNow}}
\author[ICRR]{K.~Sato}
\author[ICRR,IPMU]{H.~Sekiya}
\author[ICRR]{T.~Suzuki}
\author[ICRR]{O.~Takachio}
\author[ICRR,IPMU]{A.~Takeda}
\author[ICRR]{S.~Tasaka}
\author[ICRR,IPMU]{M.~Yamashita}
\author[ICRR,IPMU]{B.~S.~Yang\fnref{YangNow}}
\author[IBS]{N.~Y.~Kim}
\author[IBS]{Y.~D.~Kim}
\author[ISEE,KMI]{Y.~Itow}
\author[ISEE]{K.~Kanzawa}
\author[ISEE]{K.~Masuda}
\author[IPMU]{K.~Martens}
\author[IPMU]{Y.~Suzuki}
\author[IPMU]{B.~D.~Xu}
\author[Kobe]{K.~Miuchi}
\author[Kobe]{N.~Oka}
\author[Kobe,IPMU]{Y.~Takeuchi}
\author[KRISS,IBS]{Y.~H.~Kim}
\author[KRISS]{K.~B.~Lee}
\author[KRISS]{M.~K.~Lee}
\author[Miyagi]{Y.~Fukuda}
\author[Tokai1]{M.~Miyasaka}
\author[Tokai1]{K.~Nishijima}
\author[Tokushima]{K.~Fushimi}
\author[Tokushima]{G.~Kanzaki}
 \author[YNU1]{S.~Nakamura}

 \address{\rm\normalsize XMASS Collaboration$^*$}
\cortext[cor1]{{\it E-mail address:} xmass.publications11@km.icrr.u-tokyo.ac.jp .}

\address[ICRR]{Kamioka Observatory, Institute for Cosmic Ray Research, the University of Tokyo, Higashi-Mozumi, Kamioka, Hida, Gifu, 506-1205, Japan}
 \address[IBS]{Center for Underground Physics, Institute for Basic Science, 70 Yuseong-daero 1689-gil, Yuseong-gu, Daejeon, 305-811, South Korea}
\address[ISEE]{Institute for Space-Earth Environmental Research, Nagoya University, Nagoya, Aichi 464-8601, Japan}
\address[IPMU]{Kavli Institute for the Physics and Mathematics of the Universe (WPI), the University of Tokyo, Kashiwa, Chiba, 277-8582, Japan}
\address[KMI]{Kobayashi-Maskawa Institute for the Origin of Particles and the Universe, Nagoya University, Furo-cho, Chikusa-ku, Nagoya, Aichi, 464-8602, Japan}
\address[Kobe]{Department of Physics, Kobe University, Kobe, Hyogo 657-8501, Japan}
\address[KRISS]{Korea Research Institute of Standards and Science, Daejeon 305-340, South Korea}
\address[Miyagi]{Department of Physics, Miyagi University of Education, Sendai, Miyagi 980-0845, Japan}
\address[Tokai1]{Department of Physics, Tokai University, Hiratsuka, Kanagawa 259-1292, Japan}
\address[Tokushima]{Department of Physics, Tokushima University, 2-1 Minami Josanjimacho Tokushima city, Tokushima, 770-8506, Japan}
\address[YNU1]{Department of Physics, Faculty of Engineering, Yokohama National University, Yokohama, Kanagawa 240-8501, Japan}
\fntext[OgawaNow]{Now at Department of Physics, College of Science and Technology, Nihon University, Kanda, Chiyoda-ku, Tokyo 101-8308, Japan.}
\fntext[YangNow]{Now at Center for Axion and Precision Physics Research, Institute for Basic Science, Daejeon 34051, South Korea.}

 \address{   }
 \address{   }
 \address{   }
 \address{   }
 \address{   }

 \begin{abstract}
  Hidden photons and axion-like particles are candidates for cold dark matter if they were produced non-thermally in the early universe. We conducted a search for both of these bosons using 800 live-days of data from the XMASS detector with 327 kg of liquid xenon in the fiducial volume. No significant signal was observed, and thus we set constraints on the $\alpha' / \alpha$ parameter related to kinetic mixing of hidden photons and the coupling constant $g_{Ae}$ of axion-like particles in the mass range from 40 to 120 keV/$c^2$, resulting in $\alpha' / \alpha < 6 \times 10^{-26}$ and $g_{Ae} < 4 \times 10^{-13}$. These limits are the most stringent \textcolor{black}{over this mass range} derived from both direct and indirect searches to date. 
 \end{abstract}

\begin{keyword}
Dark matter \sep Hidden photon \sep Axion-like particle \sep Low background \sep Liquid xenon 

\end{keyword}

\end{frontmatter}



\section{Introduction}
\label{sec:introduction}

The existence of Dark Matter (DM) is inferred from many cosmological and astrophysical observations~\cite{PDG2016}. All indications for its existence so far were based on its gravitational interaction only, and the nature of DM beyond that is still shrouded in mystery. 
One of the theoretical realizations of DM is Weakly Interacting Massive Particles (WIMPs), which are expected to have masses of roughly 10 GeV/$c^2$ to a few TeV/$c^2$. Although many direct and indirect experiments are searching for WIMPs, no concrete signal has yet been found.
  Hidden Photons (HPs) and Axion-like Particles (ALPs), which are respectively vector and pseudo-scalar realizations of bosonic super-WIMPs~\cite{superWIMPs}, are alternative candidates for DM with expected masses $<$ 1 MeV/$c^2$.
  A HP is the gauge boson of a hidden U(1) sector that kinetically mixes with the standard model photon~\cite{HOLDOM198619j6}.
  ALPs arise as pseudo-Nambu-Goldstone bosons that appear generically in all string compactifications~\cite{wispyCDM}.
  Although a scenario in which they were thermally produced in the early universe was ruled out~\cite{XMASSCM} or disfavored~\cite{superWIMPs}, HPs and ALPs can be cold DM candidates that give rise to the observed DM abundance if they were produced non-thermally via the mis-alignment mechanism~\cite{wispyCDM}.
  HPs and ALPs are experimentally interesting because both of them are absorbed by materials with an interaction analogous to a photoelectric effect~\cite{superWIMPs}, transferring their total energy to electrons. Assuming they are cold DM, {\it i.e.} they are non-relativistic, the energy they deposit is equivalent to their rest mass. \\
      \textcolor{black}{\indent{The} calculation in Ref.\cite{superWIMPs} is limited to bosons with keV-scale mass up to roughly 100 keV/$c^2$. In this mass region, indirect searches give stringent limits of $<10^{-26}$ on the $\alpha'/\alpha$ parameter of  HP, which represents the strength of the kinetic mixing, in the mass range from 1 eV/$c^2$ to 50 keV/$c^2$ and above 140 keV/$c^2$~\cite{indirectLimit}. Around the mass of 90 keV/$c^2$, however, the indirect limit is relatively weak as $\alpha'/\alpha<O(10^{-24})$.} XMASS already carried out searches \textcolor{black}{around this mass region} for HPs and ALPs using data taken in 2010--2012, and had given limits in the mass range of 40--120 keV/$c^2$~\cite{XMASSCM}. In this paper, we report an improved result using 800 live-days of data from November 2013 to July 2016 and a fiducial volume containing 327 kg of liquid xenon. The sensitivity of the search is improved by an overall reduction of the background (BG), advances in our understanding of the BG, and a significant increase in available exposure.

\section{XMASS detector}
\label{sec:xmass-i-detector}
The XMASS detector~\cite{XMASSdetector} is located at the Kamioka Observatory, which is an underground laboratory in Japan, at a depth of 2,700-m water equivalent. The detector consists of an Inner Detector (ID) and an Outer Detector (OD).

The ID is a liquid xenon scintillator contained in a vacuum insulated copper vessel. The 832 kg of liquid xenon in the sensitive region are surrounded by 642 inward-facing photomultiplier tubes (PMTs). The PMTs are arranged in a copper holder the shape of which is an approximate sphere with a radius of $\sim$40 cm. The photocathodes of the PMTs cover 62\% of that inner surface.
Aluminum used as a seal material in the PMT contains some radioactive isotopes (RIs) which were the main BG sources in our previous analysis~\cite{XMASSCM,XMASSdetector}. In 2013 we installed copper covers over these aluminum seals to reduce this BG~\cite{XMASSFV}. Only new data taken after this installation is used in this analysis.

The OD encloses the ID, which is at its center. The OD is a water Cherenkov counter of cylindrical shape, 10 m in diameter and 10.5 m in height, and is read out by 72 20-inch PMTs. The OD is used as an active veto for cosmic-ray muons and serves as a passive radiation shield against environmental neutrons and $\gamma$-rays.

The signals from the ID-PMTs are recorded by CAEN V1751 waveform digitizers with a sampling rate of 1 GHz. Data acquisition is triggered if at least 4 PMTs detect signals within 200 ns of each other over 0.2 photoelectron (PE) threshold. 
The gains of the ID PMTs were monitored with a blue LED embedded in the inner surface of the detector, dimmed to provide single PE signal levels. The detector response to scintillation light was traced by inserting a $^{57}$Co source~\cite{XMASScalibration} into the ID every two weeks. The light yield of the detector for 122 keV $\gamma$-ray and both the absorption length and the scattering length of the liquid xenon were extracted from this regular calibration. The time evolution of these values was used as input parameters for the Geant4-based XMASS Monte Carlo simulation (MC)~\cite{XMASSdetector}.

\section{Analysis}
\label{sec:analysis}

\subsection{Event selection and vertex reconstruction}
\label{sec:event-selection}
Events without associated OD activity \textcolor{black}{(8 or more OD PMTs detect signals over a 0.4 PE threshold)} were used for the analysis.
We applied our `standard cuts', which are summarized in~\cite{XMASSFV}, to reject events originating from PMT after-pulses and electronic noise. 
We also required the timing difference to the subsequent event to be $>1$ ms, to reject $^{214}$Bi $\beta$-ray events from the $^{214}$Bi--$^{214}$Po decay sequence in the $^{222}$Rn decay chain.

\begin{figure}[tbp]
  \begin{center}
    \includegraphics[keepaspectratio=true,width=88mm]{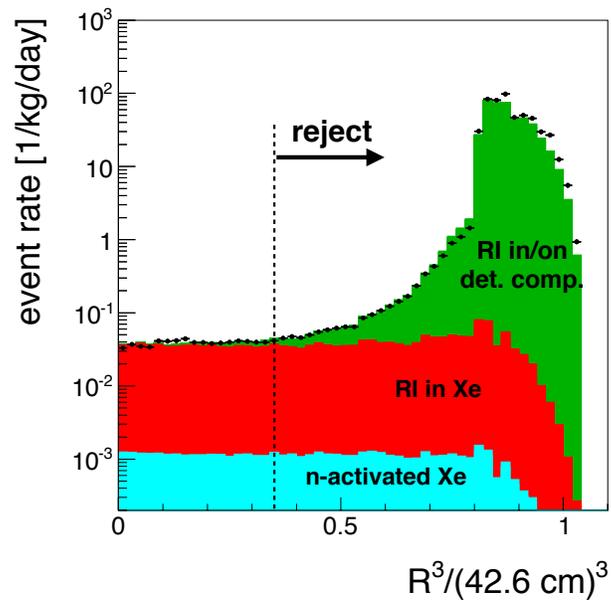}
   \caption{Distribution of reconstructed vertices as a function of radius cubed. Events passing the standard cuts with $\mathrm{NPE}_{cor}$ between 588--1764 PEs were plotted. The radius is normalized to the radius of the window surface of the most distant PMTs (42.6 cm). Black dots show the data. The stacked histograms show the BG MCs of RIs in/on the detector components (green), RIs in the liquid xenon (red), and  neutron-activated xenon isotopes (light blue). The vertical dashed line marks the cut value.}\label{fig:rcut}
  \end{center}
\end{figure}

\textcolor{black}{The event vertex was then reconstructed based on the maximum likelihood evaluation of the observed PE distribution in the ID PMTs~\cite{XMASSdetector}. The expected number of PE on each PMT was obtained from MC simulations for various vertex positions $\vec{r}$ and xenon absorption lengths $l_{abs}$, and used to calculate a likelihood for each event position. As the regular $^{57}$Co calibration provided the time evolution of $l_{abs}$, the MC expectations updated accordingly. The position $\vec{r}$ which maximizes the likelihood was accepted as the event vertex. }
%
We required the radius of the reconstructed vertex $R_{rec}$ ($\equiv |\vec{r}|$ with the center of the ID as the origin of our coordinate system) to be $<30$ cm. This fiducial volume cut strongly reduces BG events originating from $\gamma$-rays or $\beta$-rays from RI in/on the detector's inner surfaces and bulk materials; its impact is shown in Fig.~\ref{fig:rcut}.

The PE acceptance depends on the vertex position and also on optical parameters such as $l_{abs}$ and light yield in the liquid xenon. We thus use a corrected number of PEs (NPE$_{cor}$) as a measure of the energy deposit to avoid such dependencies. NPE$_{cor}$ is defined as:

\begin{equation}
 \label{eq:npecor}
 \mathrm{NPE}_{cor} \equiv \mathrm{NPE}\times\frac{\mu(\vec{r}=\vec{0},l_{abs})}{\mu(\vec{r},l_{abs})}\times\nu, 
\end{equation}
where $\mu(\vec{r},l_{abs})$ is the detection efficiency for the scintillation light generated at $\vec{r}$ when the absorption length is $l_{abs}$\textcolor{black}{, and} $\mu(\vec{r}=\vec{0},l_{abs})$ is the efficiency at the detector center. 
\textcolor{black}{These efficiencies are calculated from the MC expectations used in the vertex reconstruction.} The $\nu$ is the time dependent correction factor for the PE yield at the detector center obtained from the regular $^{57}$Co calibrations. \textcolor{black}{Changes of $\nu$ can be explained by changes in the absorption length of liquid xenon as discussed in~\cite{XMASSmodulation2017}.  The change of $\nu$ over the data taking period are controlled within $\pm$7\% from its averaged value.}

\subsection{Signal MC}

The absorption of HPs and ALPs is analogous to the photoelectric effect.
The cross-section of the absorption $\sigma_{abs}$ for HP can be written in terms of the cross-section of the photoelectric effect $\sigma_{pe}$ when replacing the photon energy $\omega$ by the HP mass $m_{HP}$, as:

\begin{equation}
 \frac{\sigma_{abs}\varv}{\sigma_{pe}(\omega =m_{HP})c}=\frac{\alpha'}{\alpha}, 
\end{equation}
where $\varv$ is the velocity of the HP, $\alpha$ is the fine structure constant, and $\alpha'$ is its analogue for the HP. Assuming a dark matter density of 0.3 GeV/cm$^3$, the event rate is expressed as~\cite{superWIMPs}:

\begin{equation}
\label{eq:rateHP}
R_{HP}[1/\textrm{kg/day}] = \frac{4 \times 10^{23}}{A}\frac{\alpha'}{\alpha} \frac{\sigma_{pe}[\textrm{barn}]}{m_{HP}[\textrm{keV}]},
\end{equation}
where $A = 131.3$ is an atomic mass of xenon of natural composition.

As for the absorption of ALPs, which is also referred to as the axio-electric effect, its cross-section and expected event rate are calculated as~\cite{superWIMPs}:
\begin{equation}
 \frac{\sigma_{abs}\varv}{\sigma_{pe}(\omega =m_{ALP})c}=\frac{3m_{ALP}^2}{4\pi\alpha f_{a}^2}, 
\end{equation}
and

\begin{equation}
\label{eq:rateALP}
R_{ALP}[1/\textrm{kg/day}] = \frac{1.2 \times 10^{19}}{A}g_{Ae}^2 \sigma_{pe}[\textrm{barn}]\cdot m_{ALP}[\textrm{keV}],
\end{equation}
where $m_{ALP}$ and $g_{Ae}$ are the mass and the axio-electric coupling constant of the ALPs, respectively. The $f_{a}$ is a dimensional coupling constant for ALPs defined as $f_{a}\equiv 2m_{e}/g_{Ae}$, where $m_{e}$ is the mass of the electron.
\textcolor{black}{In this analysis,  the expected rate does not depend on DM velocity distribution as indicated in Eqs.~(\ref{eq:rateHP}) and~(\ref{eq:rateALP}).}
In the following analysis we concentrate on HPs. The result for ALPs can be obtained by replacing the event rate from Eq.~(\ref{eq:rateHP}) by that from Eq.~(\ref{eq:rateALP}).

\begin{figure}[tbp]
  \begin{center}
    \includegraphics[keepaspectratio=true,width=88mm]{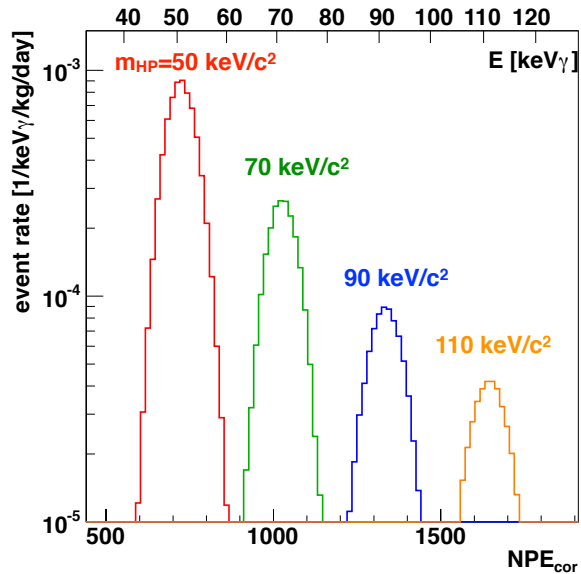}
   \caption{NPE$_{cor}$ distributions of the HP signal expected from MC with $\alpha'/\alpha = 4 \times 10^{-26}$. The corresponding $\gamma$-ray energy is also shown on the upper horizontal axis. The red, green, blue, and orange lines show signals for $m_{HP}$ of 50, 70, 90, and 110 keV/$c^2$, respectively. }\label{fig:signal}
  \end{center}
\end{figure}

We simulated the photoelectric effect of photons with energies equal to $m_{HP}$ and used this simulation as signal MC. This simulation is identical to the absorption of the HPs, because all the energy of the incident particle, including its rest mass, is transferred to electrons. \textcolor{black}{Photons were generated uniformly inside the ID. We made the momentum directions of the photons simply isotropic, since they do not affect on result of photoelectric effect. } The signal MC was made for $m_{HP}$ from 40 to 120 keV/$c^2$ in steps of 2.5 keV/$c^2$. The same selection criteria as described in Sec.~\ref{sec:event-selection} were applied.
The ratio of the number of events passing the selection to the ones generated in the 30-cm-radius fiducial volume is 0.95--0.98 for any of the HP masses.
 Examples of the expected signal spectrum are shown in Fig.~\ref{fig:signal}.

\subsection{Background MC}
\label{sec:background-mc}
MC samples were prepared for every BG source separately. The BG sources were divided into three groups: RIs dissolved in the liquid xenon, xenon isotopes activated by neutrons, and RIs in/on the detector components.

For the RIs in the liquid xenon, the amount of $^{214}$Pb in the detector, which is a daughter of $^{222}$Rn, was estimated from $^{214}$Bi--$^{214}$Po coincidences in the $^{222}$Rn decay chain to be 8.53$\pm$0.16 mBq on average throughout the data taking period~\cite{XMASS2nuECEC}. 
 The average amount of $^{85}$Kr in the detector was estimated to be 0.25$\pm$0.04 mBq using the coincidence of its $\beta$ emission with $Q_\beta =$ 173 keV followed by a 514-keV $\gamma$-ray emission~\cite{XMASSFV}. The amount of the $2\nu\beta\beta$ decay of $^{136}$Xe ($Q_{\beta\beta}=2.46$ MeV) is estimated from its natural abundance (8.9\%) assuming $T_{1/2}=2.21\times 10^{21}$ years~\cite{KamlandZen}. A contamination of argon was found from a component analysis of the detector's xenon gas. The $\beta$ decay of $^{39}$Ar can be a BG source. Also a contamination of $^{14}$C was indicated by its characteristic spectral shape in the observed energy spectrum though its chemical form is not known. The contributions of $^{39}$Ar and $^{14}$C were evaluated from the observed spectrum~\cite{XMASS2nuECEC}.

 The xenon isotopes and their daughters, mainly $^{131\textrm{m}}$Xe and $^{125}$I, are thought to be produced in gas phase by thermal neutrons when the gas is outside of the OD water shield, and then returned to the liquid xenon in the detector. The amounts of $^{131\textrm{m}}$Xe and $^{125}$I were calculated from the measured thermal neutron flux in the Kamioka Observatory $(0.8$--$1.4)\times 10^{-5}$ cm$^{-2}$s$^{-1}$~\cite{OtaniMThesis,MinaminoMThesis}.

 The amounts of RIs in/on the detector components were obtained from screening with germanium detectors and from the shape of the observed energy spectrum between 30 and 3000 keV without the fiducial volume cut, as summarized in~\cite{XMASSFV}. 

 The uncertainties of the BG amounts were considered as systematic uncertainties in our signal peak search process, as described in Sec.~\ref{sec:signal-peak-search}.

 \subsection{MC treatment and its uncertainty}
 \label{sec:discr-betw-data}

   Several corrections, which were mainly based on calibration data, were applied to the MC as summarized in Table~\ref{tab:MCdiscrepancy}. For reference purpose we label our five corrections C1 through C5. The correction C1 follows the standard MC treatment in XMASS~\cite{XMASS2nuECEC}. In addition, the corrections C2--C5 were introduced to include our knowledge of the various BG components in the energy range used in this analysis. The uncertainty in each of these corrections was used as a systematic uncertainty in the fitting process described in Sec.~\ref{sec:signal-peak-search}.

   \begin{landscape}
 \begin{table*}
  \caption{Corrections to the MC and their errors. For the non-linearity correction C1, the results of our five calibration points were connected using $f_{model}$ as described in the text. As for the energy and position resolution corrections C2 and C3, the results at $59.3$ and $122.1$ keV were interpolated with a function of energy $A \oplus B/\sqrt{E}$ where the operator $\oplus$ represents quadratic sum, and the parameters $A$ and $B$ were determined by requiring the function to connect the results at the two calibration points. }\label{tab:MCdiscrepancy}
  \begin{center}
      \small
   \begin{tabular}{c c c c c c c}
       \hline
       ID  &  & \multicolumn{5}{c}{(energy of $\gamma$-rays or events used to derive the correction)} \\
        & kind of correction & \multicolumn{5}{c}{correction factor and its error ($\equiv C_m \pm \delta C_m$ used in Eq.~(\ref{eq:chi2Sys}))}\\
        \hline 
       C1 & non-linearity of Xe scintillation& ( 5.9 keV) & ( 17.8 keV) & ( 30 keV) & ( 59.3 keV) & ( 59.5 keV) \\ 
       & $\left(\frac{\mathrm{Data}}{\mathrm{MC}}\textrm{ relative to 122.1 keV}\right)$ & $80^{+5}_{-5}$\%&$79^{+3}_{-4}$\% &$91^{+3}_{-3}$\% &$91^{+3}_{-3}$\%& $94^{+3}_{-3}$\% \\
       \hline
       C2 & NPE$_{cor}$ resolution & ( 59.3 keV) & ( 122.1 keV) \\ 
       & $\left(\sqrt{\left(\delta E/E\right)_{data}^2 - \left(\delta E/E\right)_{MC}^2 }\right)$ & $3.8\pm 2.0$\% & $1.1\pm 0.4$\%\\
       \hline
       C3 & $R_{rec}$ resolution & ( 59.3 keV) & ( 122.1 keV) \\ 
       & $\left(\sqrt{\left(\delta R_{rec}\right)_{data}^2 - \left(\delta R_{rec}\right)_{MC}^2}\right)$ & $2.6 \pm 1.1$ mm& $1.3\pm0.3$ mm \\
       \hline
       C4 & event increase due to dead PMTs&  ($441<$ NPE$_{cor}<515$)   \\ 
       & (Data/MC $-$ 1)& $(7\pm 14)$\%\\
       \hline
       C5 & event increase due to dead PMTs&  ($515 \leq$ NPE$_{cor}<588$) \\ 
       & (Data/MC $-$ 1)&  $(19\pm 16)$\%\\
       \hline 

       \end{tabular}
  \end{center}
 \end{table*}
   \end{landscape}

\begin{figure}[tbp]
  \begin{center}
    \includegraphics[keepaspectratio=true,width=88mm]{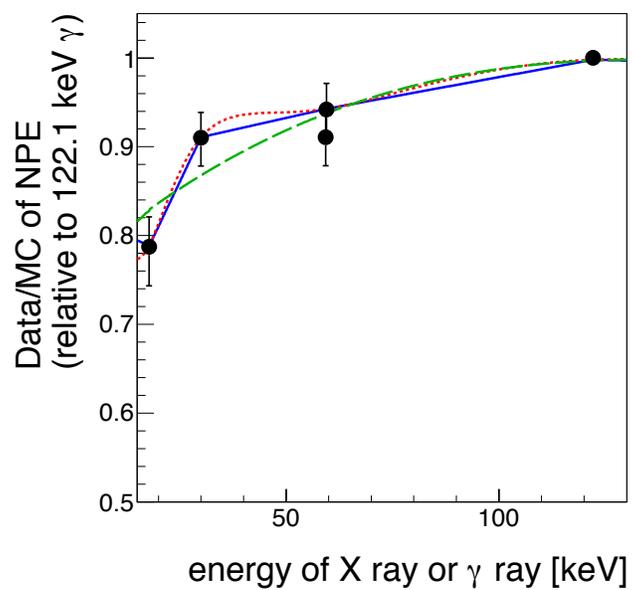}
   \caption{MC correction for the NPE non-linearity. The value at 122.1 keV $\gamma$-ray is normalized to 1. The black dots show the results from different calibration sources. Examples for the $f_{model}$ functions are shown as the solid (linear interpolation), the dotted (spline interpolation), and the dashed (6-th degree polynomial fitting) lines. } \label{fig:fmodel}
  \end{center}
\end{figure}

   The correction C1 is for the non-linear scintillation efficiency of xenon, which is important for the signal search as a change of non-linearity moves the signal peak position and deforms the BG spectrum shapes.
  The non-linearity was taken into account in our MC based on the model described in~\cite{XMASSnonlinModel}. As with the XMASS standard treatment~\cite{XMASS2nuECEC}, it was then calibrated using several $\gamma$-ray sources; $^{55}$Fe (5.9 keV), $^{241}$Am (59.5 keV $\gamma$-ray, 17.8 keV Np's X-ray, and $\sim$30 keV escape peak of Xe's X-ray), and $^{57}$Co (122.1 keV $\gamma$-ray and 59.3 keV X-ray from the tungsten housing of the source). The results were expressed relative to the 122.1 keV $^{57}$Co $\gamma$-rays. These calibration points were linearly interpolated to model the energy dependence of the non-linearity as shown in Fig.~\ref{fig:fmodel}, and the observed energy in the MC was scaled according to the model function. 
  The measurement error of each calibration point was considered as a systematic uncertainty. The uncertainty of the energy dependence model was also taken into account. In this analysis, in addition to the linear interpolation, we modeled the energy dependence with several interpolation methods (spline, and polynomial interpolation), and also fitted the calibration points with polynomial functions of various order, as well as using combinations of linear interpolation and fitting. Examples of such model functions ($f_{model}$) are shown in Fig.~\ref{fig:fmodel}. As described in Sec.~\ref{sec:signal-peak-search}, the significance of signals was evaluated including such model uncertainties.

 The resolutions for energy (NPE$_{cor}$) and position ($R_{rec}$) were evaluated from the 59.3 keV and 122.1 keV peaks in the $^{57}$Co calibration runs.
  The energy resolution in the data was evaluated to be $8.3\pm 0.8$\% at 59.3 keV and $3.9\pm 0.2$\% at 122.1 keV, which were worse than in the MC. The quadratic subtraction of the resolution in the MC from the one in the data was 3.8\% at 59.3 keV and 1.1\% at 122.1 keV.  The energy spectrum of the MC was additionally smeared to compensate for this difference (correction C2).  
  The position resolution in the data was evaluated to be $6.6\pm 0.5$ mm for 59.3 keV X-rays and $4.5\pm 0.1$ mm for 122.1 keV $\gamma$-rays at the fiducial volume radius (= 30 cm). The resolution in the MC was slightly better, the quadratic subtraction of which from the one in the data was 2.6 mm (1.3 mm) for 59.3 keV X-rays (122.1 keV $\gamma$-rays). The efficiency for the fiducial volume cut depends on the position resolution. The efficiency in the MC was thus corrected according to this difference (correction C3), but compared to the other corrections its impact on the analysis is not significant. 

  Depending on the period of data taking, eight to ten PMTs among all 642 PMTs were not operational because of their high noise rates or electrical problems. Our MC simulation shows that events generated on the detector surface near these dead PMTs are often mis-reconstructed within the fiducial volume. 
Such events come from RIs in the PMTs or $^{210}$Pb and its decay products in the copper cover for the aluminum seal, and contribute in particular to the low energy region~\cite{XMASSFV}.
  By artificially masking normal PMTs in the data, we found that the reconstructed vertices ($\vec{r}$) of such events concentrate around the axes connecting the detector center with the dead PMTs, and that reconstruction moves them towards the center of the detector.
 The probabilities of mis-reconstruction in the data and the MC were estimated using this feature~\cite{XMASSFV} for two energy regions separately. According to the difference between the probabilities, the event rate in the MC was increased by 7\% for the 441 $<$ NPE$_{cor} <$ 515 region and 19\% for the 515 $<$ NPE$_{cor} <$ 588 region (correction C4 and C5). 
 
\subsection{Signal peak search}
 \label{sec:signal-peak-search}
 We can extract a potential signal from the data by comparing the observed energy spectrum with the combined predictions of signal and BG MC including their respective uncertainties.

\begin{figure}[tbp]
  \begin{center}
    \includegraphics[keepaspectratio=true,width=88mm]{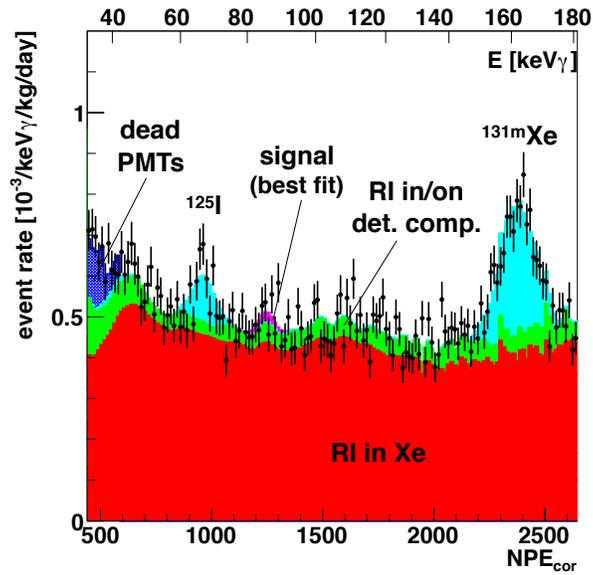}
   \caption{Best fit NPE$_{cor}$ distributions. The upper scale is translated into the corresponding $\gamma$-ray energies. The black dots represent the data . The stacked histograms show the BG MC for RIs in/on the detector components (green), RIs in the liquid xenon (red), and xenon isotopes activated by neutrons (light blue). The dark hatched blue area shows the estimated contribution from dead PMTs. The magenta part of the histogram shows the best fit HP signal for a HP mass of $m_{HP}=85$ keV/$c^2$.}\label{fig:eSpectrum}
  \end{center}
\end{figure}

  The energy spectrum after applying all the selections is shown in Fig.~\ref{fig:eSpectrum}. The peak around NPE$_{cor} = 2400$ came from residual $^{131\textrm{m}}$Xe after calibration with $^{252}$Cf, which was useful as a reference for the global energy scale of the MC.
 The energy range between NPE$_{cor}=$ 590--1760 (corresponding to $\gamma$-ray energies of 40--120 keV) was used for the signal search. In this region the spectrum is almost flat with an event rate of $\sim 5 \times 10^{-4}$ day$^{-1}$kg$^{-1}$keV$^{-1}$.

To search for the HP signal, the histogram of the observed energy spectrum was fitted with the sum of a putative signal and the BG MC spectra.
The range of the histogram used for fitting was 440--2650 NPE$_{cor}$ (corresponding to $\sim$ 30--180 keV $\gamma$-ray energy), and divided equally into 150 NPE$_{cor}$ bins.
 The chi-square was defined as:

\begin{equation}
 \label{eq:chi2}
    \chi^2_{fit}\left(m_{HP},\alpha'/\alpha\right)\equiv\sum_{i}^{N_{bin}} \frac{\left( R_{obs}^{i} - R_{BGtot}^{i} -R^{i}_{HP}(m_{HP},\alpha'/\alpha)\right)^2}{\left(\delta R_{obs}^{i}\right)^2+\left(\delta R_{BGtot}^i\right)^2+\left(\delta R_{HP}^i\right)^2} + \chi^2_{sys},
\end{equation}
where $R_{obs}^{i},R_{BGtot}^{i}$, and $R^{i}_{HP}$ are the event rate in the $i$-th bin for data, BG MC, and HP MC, respectively, and the $\delta R_{obs}^i$, $\delta R_{BGtot}^i$, and $\delta R_{HP}^i$ are their respective statistical errors. $R_{BGtot}$ is the sum of the different BG MC event rates, {\it i.e.}:

\begin{equation}
\label{eq:RBG}
 R_{BGtot} = \sum_{j:\textrm{RI types}}p_{j}R_{j\textrm{-th BG}},
\end{equation}
where the summation is taken for all the RIs of the three BG categories described in Sec.~\ref{sec:background-mc}. The $R_{j\textrm{-th BG}}$ is the expected event rate from the $j$-th RI and $p_j$ is its scale parameter whose initial value is unity.
The $\chi^2_{sys}$ is a penalty term to handle systematic uncertainties. It is defined as:

\begin{equation}
\label{eq:chi2Sys}
 \chi^2_{sys} = \sum_{j:\textrm{RI types}}^{j\neq ^{14}C,^{39}Ar}\left(\frac{1-p_{j}}{\delta p_j}\right)^2 + \sum_{m=1}^{5}\left(\frac{\Delta C_m}{\delta C_m}\right)^2,
\end{equation}
where the $\delta p_j$ are the $1\sigma$ uncertainties of the BG estimates described in Sec.~\ref{sec:background-mc}. The first summation on the right side constrains the parameters $p_j$ around $\pm 1 \sigma$ from unity during the fit. Because the amounts of $^{14}$C and $^{39}$Ar in xenon are obtained directly from the fit to the data, these RIs are not included in this summation while they are included in the summation in Eq.~(\ref{eq:RBG}).
 The second summation is a penalty term related to the five special MC corrections C1--C5 described in Sec.~\ref{sec:discr-betw-data}. The $m$-th correction factor $C_m$ listed in Table~\ref{tab:MCdiscrepancy} is modified by $\Delta C_m$ in the fit, and its uncertainty $\delta C_m$ constrains this modification.
 $R_{j\textrm{-th BG}}$ and $R_{HP}$ are functions of the $\Delta C_m$, the model function type $f_{model}$ described in Sec.~\ref{sec:discr-betw-data}, and a global energy scale $\epsilon_{E}$:

 \begin{equation}
 R_{j\textrm{-th BG}} = R_{j\textrm{-th BG}}\left(\epsilon_{E},\Delta C_1,...,\Delta C_5;f_{model}\right),
 \end{equation}
\begin{equation}
  R_{HP} = R_{HP}\left(m_{HP},g_{Ae};\epsilon_{E},\Delta C_1,...,\Delta C_5;f_{model}\right).
\end{equation}
 The $\chi^2_{fit}$ was minimized separately for every 2.5 keV/$c^2$ step in HP mass between 40 and 120 keV/$c^2$ and for each step of $\alpha'/\alpha$ in the $\alpha'/\alpha>0$ region, by fitting the $p_j$, $\Delta C_m$, $\epsilon_{E}$, and $f_{model}$. The optimization of these parameters except for $f_{model}$ was done using ROOT TMinuit~\cite{ROOT}. As for $f_{model}$, each $f_{model}$ was tested separately, and the one giving the smallest $\chi^2_{fit}$ was chosen for each mass and each $\alpha'/\alpha$.
 This method corresponds to handling the model function shape as one of the fitting parameters~\cite{descreteProfile}. The sensitivity obtained with this method is conservative compared to sticking with one model. We thus obtained the $\chi^2_{fit}$ profile as a function of $\alpha'/\alpha$ for each mass. The minimum of the profile is the most probable $\alpha'/\alpha$ parameter for that mass.

\section{Result}
\label{sec:result}

\begin{figure}[tb]
  \begin{center}
   \includegraphics[keepaspectratio=true,width=88mm]{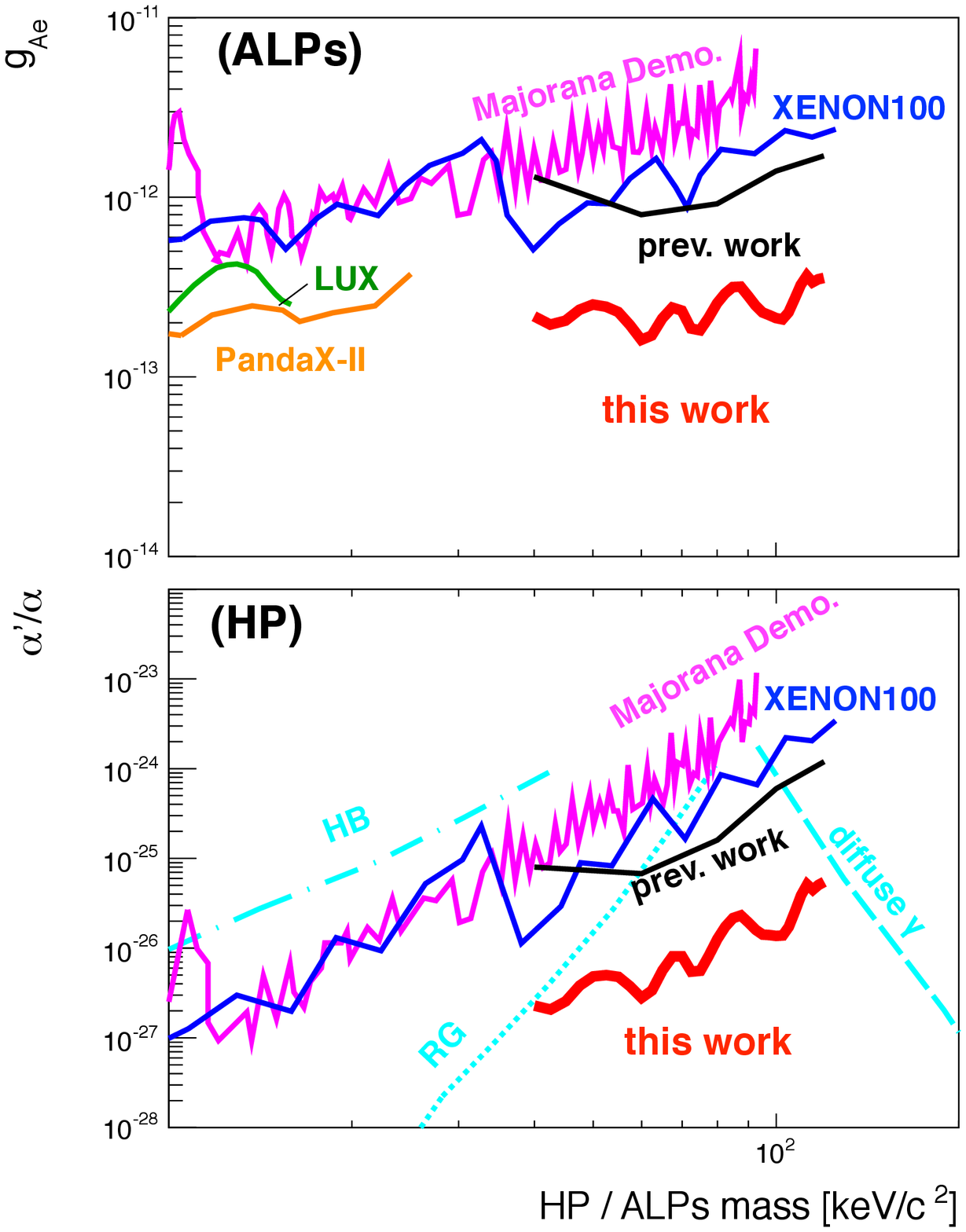}
   \caption{Constraints on $g_{Ae}$ of the ALPs (top) and $\alpha'/\alpha$ of the HP (bottom). The red line shows the 90\%CL constraint presented in this paper. The black line shows our previous result~\cite{XMASSCM}. The blue, magenta, green, and orange lines are limits reported by the XENON100~\cite{XENON100}, the Majorana Demonstrator~\cite{Majorana}, the LUX~\cite{LUX}, and the PandaX-II~\cite{PandaX}. The dotted, dashed, and dash-dotted lines in light blue color are constraints from indirect searches derived from red giant stars (RG), diffuse $\gamma$-ray flux, and horizontal branch stars (HB), respectively~\cite{indirectLimit}.}\label{fig:compare}
  \end{center}
\end{figure}

 The best fit result for HP with $m_{HP}=85$ keV/$c^2$ is shown in Fig.~\ref{fig:eSpectrum}, where the minimum $\chi^2_{fit}$/NDF = 131/122 with $\alpha'/\alpha = 1.1 \times 10^{-26}$. \textcolor{black}{No significant signal was found at any HP mass.} 
 The difference between the minimum $\chi^2_{fit}$ and the $\chi^2_{fit}$ with no signal was at most 1.62. 
 {\color{black} We thus set the 90\% confidence level (CL) constraint on $\alpha'/\alpha$ from the relation:
 
 \begin{equation}
  \frac{\int_0^{a_{90}}\exp\left(-\chi^2_{fit}/2\right)da}{\int_0^{\infty}\exp\left(-\chi^2_{fit}/2\right)da} = 0.9,
 \end{equation}
 where $a$ and $a_{90}$ denote $\alpha'/\alpha$ and its constraint, respectively.}
  The constraint for each mass is shown in Fig.~\ref{fig:compare}. Compared to our previous work, the constraints improved by a factor of 10--50. The constraints from other direct and indirect searches are also shown in the figure.  Our result gives the most stringent limit in the mass range from 40 to 120 keV/$c^2$. The indirect limits for HP around 90 keV/$c^2$ are $\alpha' / \alpha  < O(10^{-24})$, relatively weak compared to the higher and lower mass regions ($\alpha' / \alpha  <  O(10^{-27})$ for $m_{HP}\geq 200$ or $\leq 40$ keV/$c^2$). Our limit, \textcolor{black}{$\alpha' / \alpha \leq 2\times 10^{-27}$}--$6\times 10^{-26}$, bridges that region of relative weakness for the indirect limits.  

  The same result converted to a constraint on $g_{Ae}$ for ALPs using Eqs.~(\ref{eq:rateHP})~and~(\ref{eq:rateALP}) is also shown in the figure. LUX and PandaX-II present limits for ALPs below 20 keV/$c^2$. Our result covers the higher mass region. While XENON100 and the Majorana Demonstrator report limits over a wider mass region, our constraint $g_{Ae} < $ a few $10^{-13}$ is the best limit for $m_{ALP} > 40 $ keV/$c^2$.
 
\section{Conclusion}
\label{sec:conclusion}

The searches for HPs and ALPs, which are candidates for cold DM, were conducted using 800 live-days XMASS data with 327 kg xenon in a 30-cm-radius fiducial volume. We searched for signal peaks from HPs or ALPs interactions analogous to the photoelectric effect in the energy spectrum around tens of keV, where the event rate in XMASS is $\sim 5 \times 10^{-4}$ day$^{-1}$kg$^{-1}$keV$^{-1}$. With the absence of any significant signal, we set the most stringent upper limits for the parameter for kinetic mixing $\alpha'/\alpha$ of the HP and for the coupling constant $g_{Ae}$ of the ALPs  in the mass range from 40 to 120 keV/$c^2$.

\section*{Acknowledgments}
We gratefully acknowledge the cooperation of the Kamioka Mining and Smelting Company.
This work was supported by the Japanese Ministry of Education,
Culture, Sports, Science and Technology, Grant-in-Aid for Scientific Research, 
JSPS KAKENHI Grant No. 19GS0204 and 26104004,
the joint research program of the Institute for Cosmic Ray Research (ICRR), the University of Tokyo, and partially by the National Research Foundation of Korea Grant (NRF-2011-220-C00006) \textcolor{black}{and Institute for Basic Science (IBS-R017-G1-2018-a00).}


\bibliographystyle{apsrev4-1}

\bibliography{reference}
\end{document}